# Dependence of Maximum Trappable Field on Superconducting Nb$_3$Sn Cylinder Wall Thickness*

**Mario Rabinowitz**
*Electric Power Research Institute, Palo Alto, California 94303*

**H. W. Arrowsmith and S. D. Dah1gren**
*Battelle-Northwest, Richland, Washington 99352*

Inquiries to: *Armor Research*
*715 Lakemead Way, Redwood City, CA 94062*
Mario715@earthlink.net

**Abstract**

Uniform dipole magnetic fields from 1.9 to 22.4 kOe were permanently trapped, with high fidelity to the original field, transversely to the axes of hollow Nb3Sn superconducting cylinders. These cylinders were constructed by helically wrapping multiple layers of superconducting ribbon around a mandrel. This is the highest field yet trapped, the first time trapping has been reported in such helically wound taped cylinders, and the first time the maximum trappable field has been experimentally determined as a function of cylinder wall thickness.

Uniform dipole magnetic fields were permanently trapped transversely to the axes of hollow Nb3Sn superconducting cylinders, and the dependence of the maximum trappable field on superconductor thickness was investigated. This is the first time magnetic fields have been trapped in such helically wound taped cylinders, as well as the first time the maximum trappable field has been experimentally determined as a function of wall thickness. The highest field trapped was 22.4 kOe, which is higher than the highest trapped field of 17. 6 kOe reported so far.[1]  Rabinowitz [1], and Rabinowitz, Garwin and Frankel [2,3] have studied the trapping of dipole, quadrupole, and sextupole magnetic fields in hollow superconducting cylinders, and Smith [4] has reported theoretical work on flux trapping. A related study was conducted by Martin and St. Lorant [5], who investigated the shielding of magnetic fields with a superconducting flux exclusion tube.

Magnetic fields were trapped in superconducting tubes by two procedures. [1-3] In procedure I, the external dipole field was turned on and the hollow superconducting cylinder then cooled to below its transition temperature in the external field. Next, the external field was turned off and the field remained trapped in the superconductor with high fidelity to the original field. In procedure II, the superconductor was first cooled below its transition temperature. Then the external field was turned on to drive the magnetic field into the superconducting cylinder by exceeding the shielding limit. Finally the external field was turned off and the field remained stored in the cylinder. A schematic of the field trapped in a cylinder and the experimental arrangement is reproduced in Fig. 1.

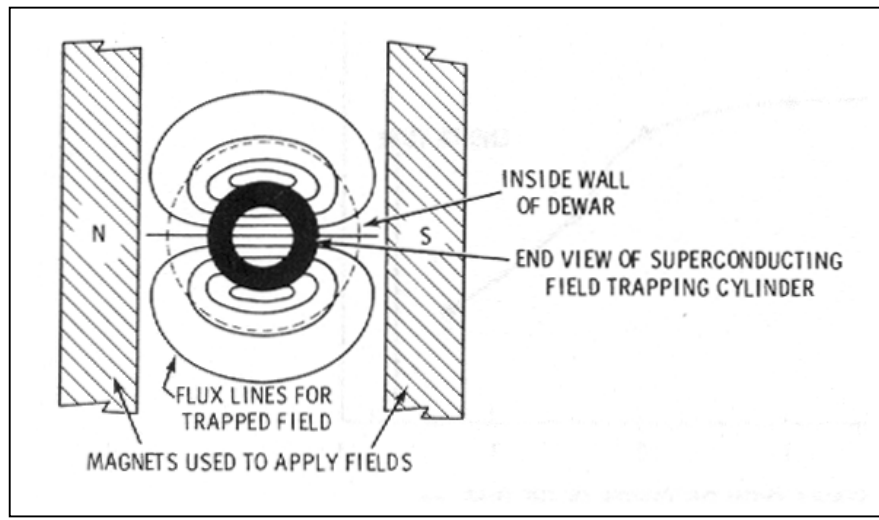

FIG. 1. Experimental arrangement used and end view schematic of the field trapped in a Nb3Sn superconducting cylinder after the external field has been turned off.

Superconducting cylinders 7.62 cm long used for the field trapping experiments were helically wound from Nb3Sn ribbon 2.54 cm wide by 75 µm thick on nominally

2.54 cm diameter mandrels. The thickness of the superconductor in the wall of the cylinder was determined by the number of layers of ribbon that were applied; cylinders containing 2, 10, 20, and 44 layers were used for the experiments. Alternate layers were helically wound in the opposite direction so that the short piece of ribbon in one layer diagonally crossed the ribbon piece beneath it. The cylinders could just as well have been constructed by placing pieces of ribbon parallel to the axis of the cylinder, but they were helically wound for convenience in construction. A solenoidal configuration was not created by the helical geometry because alternate pieces of ribbon were wound in opposite directions and there were no superconducting electrical connections at the cylinder ends. The Cu-clad ribbon (purchased from IGC) consisted of a center layer of unreacted Nb, a layer of 10.2 µm thick $Nb_3Sn$ on each side of the Nb layer, and a thin outside layer of Sn on the $Nb_3Sn$. The critical current density of the ribbon was 9.6 x $10^4$ A/$cm^2$ at 100 kOe, and we measured $T_c$ to be 16.7 - 17.0 K. The wound cylinders were heated to 240 °C after wrapping to melt the Sn on the ribbon surface and thus bond the ribbon layers to each other so that they would not move during the field trapping experiments. A movable Hall probe was used to measure the field at the center of the cylinder.

TABLE I. Summary of experimental results.

| No. of ribbon layers | Superconductor thickness in cylinder wall (mm) | Test No. | External applied field (kOe) | Trapping procedure | Trapped field (kOe) |
|---|---|---|---|---|---|
| 2 | 0.04 | 1 | 3.2 | I | 1.9 |
| 10 | 0.2 | 2 | 30 | I | 8.8 |
| 20 | 0.4 | 3 | 20 | I | 14.7 |
| | | 4 | 40 | II [a] | 15.2 |
| | | 5 | 41.5 | II | 15.3 |
| 44 | 0.9 | 6 | 30 | I | 22.4 |
| | | 7 | 36 | I | 22.4 |
| | | 8 | 51 | II | 21.2 |

[a] Procedure II was used for test No. 4 on the cylinder that already had a field of 14.7 kOe trapped in it from test No. 3 using procedure I.

The field that could be trapped increased from 1.9 to 22.4 kOe as the superconductor thickness in the wall of the tube increased from 0.04 mm (2 ribbon layers) to 0.9 mm (44 layers, Table I and Fig. 2). Moreover, the maximum field that could be trapped was rather insensitive to the field applied to the tube or to the procedure used to trap the field (Table I). For example, 22.4 kOe was trapped in the 44-layer tube with procedure I when the applied field was 30 kOe, which was the same field trapped when 36 kOe was applied (Table 1). Only slightly less field was stored (21.2 kOe) when 51 kOe was applied using procedure II (Table I and Fig. 3). Similarly, approximately 15.2 kOe was trapped in the 20-ribbon-layer cylinder with 40 kOe applied using procedure II, whereas only 0.5 kOe less (14.7 kOe) was trapped in the same cylinder with 30 kOe using procedure I (Table I). We observed, perhaps significantly, that when using procedure II, the applied field (~22 kOe) at which a field first penetrated the cylinder (i. e., the shielding limit) was almost the same as the magnitude of the field that could be trapped (21.2 kOe, Fig. 3).

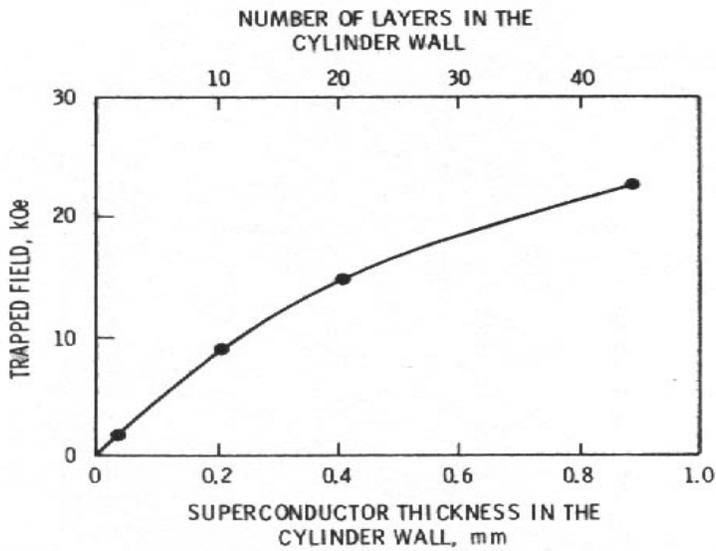

FIG. 2. Dependence of maximum trappable field on the Nb3Sn superconductor thickness. The fields were trapped using procedureI.

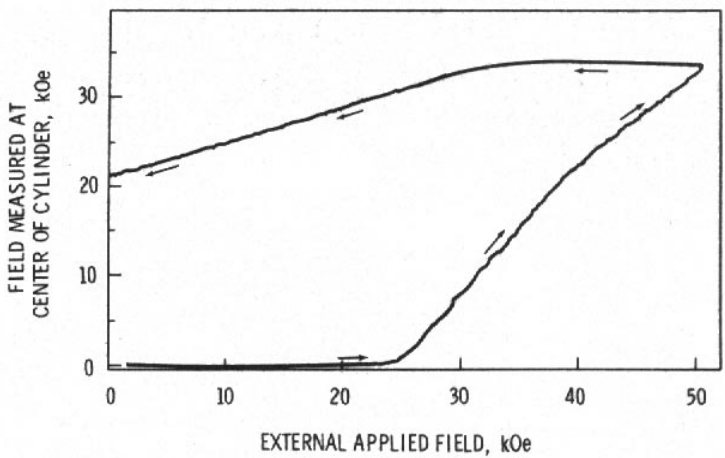

FIG. 3. Trapping 21.2 kOe in the 44-ribbon-layer Nb3Sn cylinder using procedure 11. Field penetration and ultimately trapped field versus applied field.

The applied and subsequently trapped dipole fields were measured similarly to the method previously reported [2, 3]; this showed that the shape of the applied field was faithfully reproduced within the accuracy of the Hall probe measurements. Measurements of the trapped field along the length of the 44-layer tube showed that the field was uniformly trapped at 22.4 kOe over the central 2 cm of the tube and was still within 10% of the maximum field (22.4 - 20.2 kOe) over the central 4 cm of the 7. 62-cm-long tube (Fig. 4). The trapped field decreased rapidly outside the central 4 cm of the tube (Fig. 4).

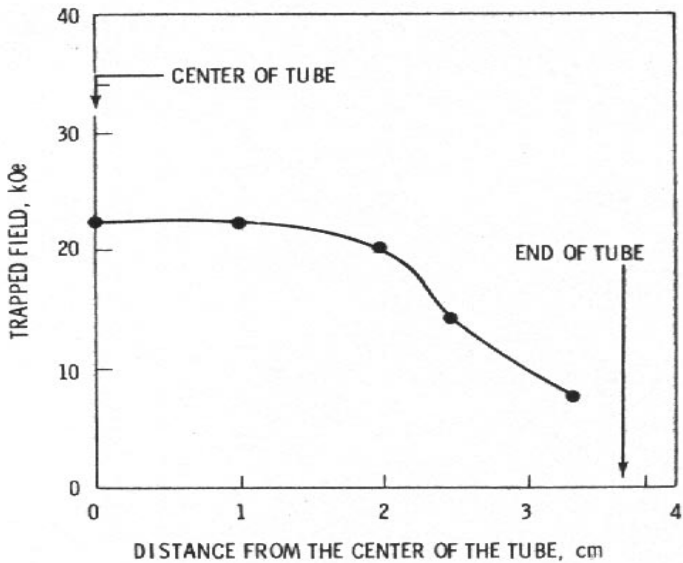

FIG. 4. Profile of the trapped dipole field near the cylinder centerline from the center of the cylinder to near one end for the cylinder with 44 wraps of Nb3Sn superconductor.

The trapped fields were stable and did not decay from the superconducting cylinders over the time available to test the time stability of the trapping. Our limited time at the magnet facility necessitated releasing the trapped field after times on the order of 1 h; nevertheless, Rabinowitz [1] reported that fields were stably stored for 1 1/2 days and predicted they would be stable for much longer periods of time.

Information on the field trapping mechanism was not obtained in the present experiments, but Rabinowitz [1] discussed two mechanisms to explain the results of previous similar experiments on superconducting field trapping tubes. It was suggested [1] that the fields are maintained either by the circulation of macroscopic currents or by microscopic vortex currents whose vector sum is equivalent to macroscopic currents. The latter view is favored although it is not necessary to rely on the microscopic current mechanism to explain the trapping of fields in cylinders made up of discontinuous pieces of superconducting ribbon.

The 22.4 kOe trapped in the 44-ribbon-layer cylinder is the highest field yet trapped in a superconducting tube. It appears that still higher fields can be trapped in cylinders with thicker superconducting walls because the field that can be trapped in the cylinders is clearly dependent upon the wall thickness as shown by these experiments. Moreover, the amount of field that can be trapped is expected to be dependent upon the superconducting properties of the material used. For example, superconductors with higher critical current density (i. e., with high pinning force) are expected to trap higher fields. Thus studies of the relationship between pinning force and field trapping ability are planned for a number of materials. We also plan to determine trapping in a single solid layer of superconductor in the walls of the cylinder relative to a cylinder fabricated from the same thickness of multiple superconducting layers. The former will have much less surface pinning.

The advantages of storing magnetic fields in superconducting tubes are significant, For example, fields can be reproduced many times from a pattern magnet and used in place of the magnet. Thus only one expensive pattern magnet is required. The stored fields also are expected to be extremely stable and thus should be valuable for many electric utility applications such as in circuit breakers, generators and magnetic refrigerators, for MHD and controlled thermonuclear fusion, as well as for focusing charged particles for accelerators or in electron microscopes.

The authors acknowledge Steve Faber's contribution to the experimental work. Use of the Brookhaven National Laboratory dipole magnet and the assistance at BNL of Bill Sampson is gratefully acknowledged.

*Work supported by the Electric Power Research Institute (EPRI).